\documentclass[3p,times]{elsarticle}

\usepackage{ecrc}
\usepackage{user}

\volume{00}
\firstpage{1}

\journalname{Nuclear Physics A}

\runauth{Xiaoming Zhang for the ALICE Collaboration}

\jid{nupha}
\jnltitlelogo{Nuclear Physics A}

\begin{document}

  \begin{frontmatter}

  \title{$\Kshort$ and $\Lambda$ Production in Charged Particle Jets in p--Pb
         Collisions at $\sNN=5.02$~TeV with ALICE}

  \author{Xiaoming Zhang (for the ALICE\fnref{colA} Collaboration)}
  \fntext[colA]{A list of members of the ALICE Collaboration and
                acknowledgements can be found at the end of this issue.}
  \address{Lawrence Berkeley National Laboratory,
           1 Cyclotron Road, Berkeley, CA 94720, U.S.A.}

  \begin{abstract}
  We study the production of $\Kshort$ mesons and $\Lambda$ baryons in jets
  in p--Pb collisions at $\sNN=5.02$~TeV with ALICE at the LHC.
  The $\pT$-differential density of the particles produced in jets is
  compared to the inclusive distributions and the $\Lambda/\Kshort$ ratio
  is reported in bins of multiplicity of the collisions.
  The hard scatterings are selected on an event-by-event basis using
  the anti-$k_{\rm T}$ clustering algorithm with resolution
  parameter $R=0.2,~0.3$ and $0.4$,
  reconstructed from charged particles with a minimum $p_{\rm T,jet}$
  of $10$ (or $20$) $\GeVc$.
  \end{abstract}


  \end{frontmatter}

  \section{Introduction}\label{sec:c01Intro}

  The strong suppression of high $\pT$ charged
  particles~\cite{Aamodt:2011qy,Abelev:2012hxa} and jets~\cite{Abelev:2013kqa}
  measured in heavy-ion collisions at the LHC is not present in p--Pb
  collisions at $\sNN=5.02$~TeV~\cite{Abelev:2014dsa}.
  However, the "double-ridge" structure observed in the long-range
  two-particle correlations in high multiplicity
  p--Pb collisions~\cite{Abelev:2012ola,ABELEV:2013wsa} resembles
  collective features found in Pb--Pb collisions~\cite{Aamodt:2010uf}.
  It has been suggested that some final state effects,
  such as parton-induced interactions~\cite{Strikman:2011cx} and strong
  correlations in particle production characteristic of a high density
  system~\cite{Bozek:2013uha,Shuryak:2013ke} may be
  present in p--Pb collisions.
  In addition, an enhancement of the baryon-to-meson yield ratio at
  intermediate $\pT\sim 3~\GeVc$ in high multiplicity p--Pb collisions
  as compared to pp collisions~\cite{Abelev:2013haa} has been observed.
  This effect is qualitatively similar to the enhancement observed
  in Pb--Pb collisions~\cite{Abelev:2013xaa} that has been discussed in
  terms of collective flow~\cite{Bozek:2011gq},
  which could be presented in small systems like pp
  collisions~\cite{Ortiz:2013yxa},
  and/or parton recombination~\cite{Fries:2003vb}.
  To discriminate between hard and soft processes contributing to the
  baryon and meson production,
  we study the $\Lambda/\Kshort$ ratio within jets and compare it to
  inclusive distributions measured in p--Pb and minimum-bias simulations
  generated by PYTHIA8 (tune 4C)~\cite{Sjostrand:2007gs} for
  pp collisions at $\s=5.02$~TeV. 

  \section{Experimental Setup and Analysis Strategy}\label{sec:c02AnaLoop}

  This analysis is performed for minimum bias p--Pb events at $\sNN=5.02$~TeV,
  corresponding to integrated luminosity
  of $\mathcal{L}_{\rm int}=51~\mu{\rm b}$.
  For a detailed description of the ALICE detector and its performance,
  see ~\cite{Aamodt:2008zz,Abelev:2014ffa}.
  The event multiplicity classes are determined by the VZERO-A detector (V0A)
  covering pseudorapidity of $2.8<\eta<5.1$ in the Pb-going direction.
  The uncertainty of multiplicity estimation is made using energy deposition
  in the Pb-going side Zero Degree Calorimeter (ZNA) and determined by
  the so-called {\it hybrid method}~\cite{Toia:2014cd}.

  Charged particles are measured using the Inner Tracking System (ITS) and
  the Time Projection Chamber (TPC).
  The ITS also provides measurement of the primary vertex.
  The selection of tracks used for the jet
  reconstruction (see~\cite{Abelev:2013fn} for details) ensures an almost
  uniform tracking efficiency in full azimuth.
  Tracks with $\pT>150$~MeV/$c$ are retained in the analysis.
  Charged particle jets are reconstructed with the
  anti-$k_{\rm T}$ algorithm~\cite{Cacciari:2008gp},
  with resolution parameter $R=0.2,~0.3$ and $0.4$.
  To remove combinatorial jets a cut on the jet
  area $A_{\rm jet}$~\cite{Cacciari:2007fd} is applied and only jets
  with $A_{\rm jet}>0.6\pi R^{2}$ are retained.
  The reconstructed jet $\pT$ is obtained by subtracting the energy of
  the underlying background according to the approach
  introduced in~\cite{Cacciari:2007fd}:
  $p_{\rm T,jet}^{\rm ch}=p_{\rm T,jet}^{\rm det}-\rho\times A_{\rm jet}$,
  where $p_{\rm T,jet}^{\rm det}$ is the measured jet $\pT$,
  and $\rho$ is the underlying background density obtained with the median
  occupancy method~\cite{Chatrchyan:2012tt}.
  The analysis is performed separately for two selections of jets:
  with $p_{\rm T,jet}^{\rm ch}>10~\GeVc$ and
  with $p_{\rm T,jet}^{\rm ch}>20~\GeVc$.
  A fiducial cut was applied requiring the jet centroid to be within the
  detector acceptance $|\eta_{\rm jet}|<0.75-R$.

  The $\Lambda$ and $\Kshort$ ($\Vzeros$) are reconstructed via the
  hadronic decays,
  $\Lambda\to{\rm p}\pi^{-}$ and $\Kshort\to\pi^{+}\pi^{-}$.
  Decay daughters are identified by their specific ionization,
  ${\rm d}E/{\rm d}x$, in the TPC.
  The yield of $\Vzero$ signal is extracted using an invariant mass analysis
  with the bin counting method~\cite{Chinellato:2012lf}.
  Tracks used for jet reconstruction are required to point to the
  primary vertex.
  Since this selection removes tracks from secondary weak decays,
  the $\Vzero$ reconstruction is performed independently of the
  jet reconstruction.
  To obtain the yield of $\Vzero$ particles associated with jets,
  each $\Vzero$ particle is matched with a jet if the distance within
  the ($\eta-\varphi$) plane between the particle and the jet axis,
  $R(\Vzero,{\rm jet})$ is smaller than the resolution parameter $R$: 
  $R(\Vzero,{\rm jet})<R$.
  In addition, $\Vzero$ particles were required to fulfill a fiducial cut
  of $|\eta_{\rm\Vzero}|<0.75$. 

  The $\Vzero$ particles not associated to a hard scattering are
  selected using two criteria:
  \begin{itemize}
  \item {\it outside cone} (OC $\Vzero$):
        the $\Vzero$ particles reconstructed outside the jet cone of any of
        the reconstructed jets in the event,
        $i.~e.$~$R(\Vzero,{\rm jet})>R_{\rm cut}$,
        where $R_{\rm cut}=0.4,~0.6$ and $0.8$;
  \item {\it non-jet events} (NJ $\Vzero$):
        $\Vzero$ particles found in events without any jet
        having $\pT>5~\GeVc$.
  \end{itemize}
  The uncertainty on the contribution of background $\Vzero$ particles to
  the jet signal is given by the difference between the spectra of
  NJ $\Vzeros$ and OC $\Vzeros$ with various $R_{\rm cut}$ values.

  The reconstruction efficiency of the $\Vzero$ particles was
  obtained (separately for $\Kshort$ and $\Lambda$) by a detector-level
  simulations incorporating a realistic detector description,
  using the GEANT3 package.
  Since the efficiency of the $\Vzero$ depends on pseudorapidity and is
  sensitive to the $\eta$ dependent distribution of jets,
  the efficiency obtained from simulations was weighted with the jet
  distribution obtained in the data.

  The per event $\pT$-differential density ${\rm d}\rho/{\rm d}\pT$ of $\Vzero$
  particles is given by the corrected $\pT$ spectrum,
  ${\rm d}N/{\rm d}\pT$ scaled by the area of the acceptance given either by
  the jet area or the acceptance not associated to the jet
  production (OC or NJ selection):
  \begin{equation}
  \frac{{\rm d}\rho}{{\rm d}\pT}=\frac{1}{N_{\rm ev}}\times
  \frac{1}{\langle{\rm Area}\rangle}\times
  \frac{{\rm d}N}{{\rm d}\pT},
  \end{equation}
  where $N_{\rm ev}$ is the number of events and $\langle{\rm Area}\rangle$
  is the averaged per event acceptance
  of $\Vzeros$ ($\Delta\eta\times\Delta\varphi$).
  The $\pT$-differential density of $\Vzero$ particles produced in jets is
  obtained by subtracting the underlying $\Vzero$ density from the $\Vzeros$
  matched with jets.
  Feed-down of $\Lambda$ from $\Xi$ decays in jets is corrected using the
  feed-down fraction of inclusive $\Lambda$ obtained from data.
  The systematic uncertainty of the fraction of feed-down of $\Lambda$ in jets
  was estimated using the difference between the PYTHIA8 simulation and the
  inclusive measurements at high-$\pT$ in the data.
  The uncertainty on jet background fluctuations is evaluated by varying the
  jet $\pT$ threshold by $20\%$.

  \section{Results}\label{sec:c03Results}

  \begin{figure}[!h]
  \begin{center}
  \includegraphics[width=.6\textwidth]{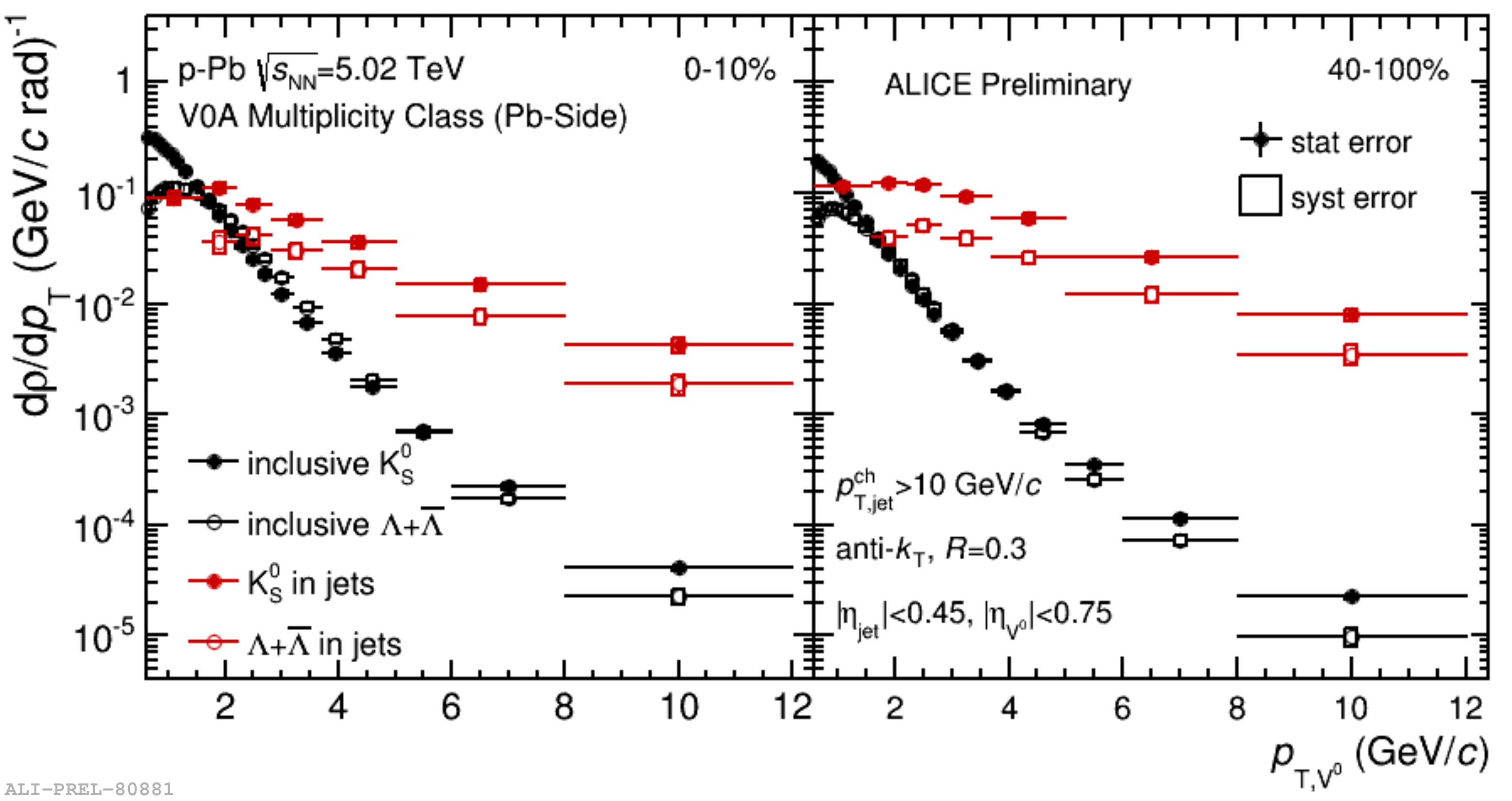}
  \caption{The $\pT$-differential density of $\Kshort$
           and $\Lambda$ ($\AntiLa$) in jets
           with $p_{\rm T,jet}^{\rm ch}>10~\GeVc$ in p--Pb collisions.
           Jets are reconstructed using charged particles and
           the anti-$k_{\rm T}$ algorithm with $R=0.3$.
           Results are shown for two V0A multiplicity classes of
           p--Pb collisions at $\sNN=5.02$~TeV: $0-10\%$ (left),
           and $40-100\%$ (right).
           Results are compared to the inclusive $\Vzero$ $\pT$ distribution.}
  \label{fig:V0dens}
  \end{center}
  \end{figure}

  \begin{figure}[!h]
  \begin{center}
  \includegraphics[width=.6\textwidth]{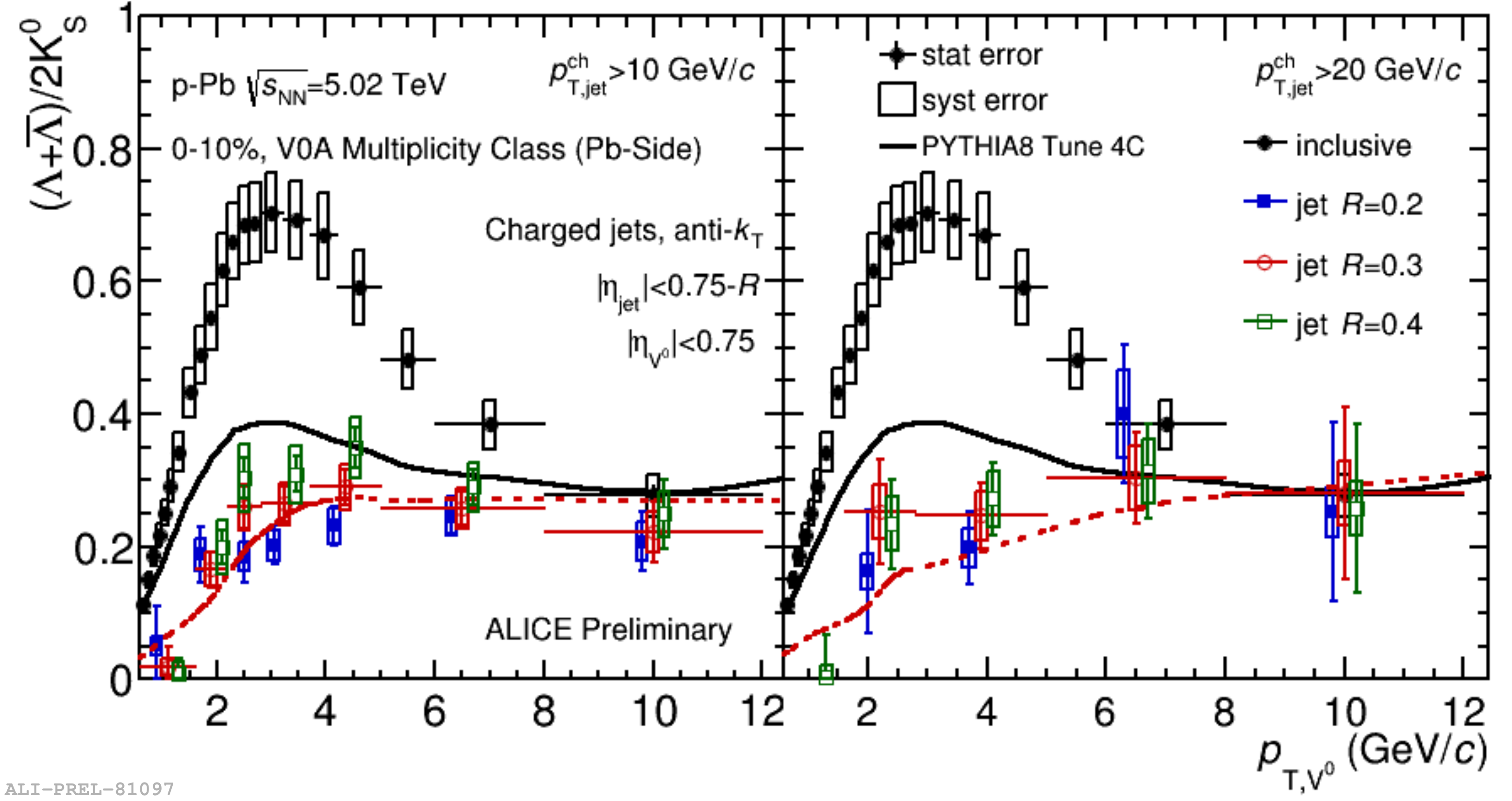}
  \caption{$\Lambda/\Kshort$ ratio in jets
           with $\pT^{\rm jet}>10$~GeV/$c$ (left)
           and $\pT^{\rm jet}>20$~GeV/$c$ (right) for events in the $0-10\%$
           V0A multiplicity class of p--Pb collisions at $\sNN=5.02$~TeV.
           Jets are reconstructed using charged particles and the
           anti-$k_{\rm T}$ algorithm.
           Results are shown for different jet $R$ of 0.2, 0.3, and 0.4.
           The ratios in jets are compared to the inclusive distributions
           and the corresponding PYTHIA8 simulations (the pp collisions).}
  \label{fig:V0ratio}
  \end{center}
  \end{figure}

  Fig.~\ref{fig:V0dens} shows $\pT$-differential density of $\Vzeros$ produced
  in jets (red points) after the underlying background subtraction and
  feed-down correction in $0-10\%$ (left) and $40-100\%$ (right)
  event multiplicity classes.
  The results are obtained for charged particle jets with $R=0.3$
  and $p_{\rm T,jet}^{\rm ch}>10~\GeVc$.
  A much harder spectrum is observed for the $\Vzeros$ produced in jets,
  relative to the inclusive $\Vzeros$ (black points).

  The $\Lambda/\Kshort$ ratios in jets with $R=0.2,~0.3$ and $0.4$ are
  presented in Fig.~\ref{fig:V0ratio}.
  The results are obtained in the $0-10\%$ event multiplicity class
  with $p_{\rm T,jet}^{\rm ch}>10~\GeVc$ and $>20~\GeVc$ and compared to the
  inclusive $\Vzero$ in data and in PYTHIA8~\cite{Sjostrand:2007gs} simulations.
  The ratios in jets do not depend on the jet radii and do not vary
  with $p_{\rm T,jet}$.
  At the intermediate $\pT$ ($2<\pT<4~\GeVc$) the inclusive ratio in p--Pb
  collisions is much larger than in PYTHIA-generated pp collisions.
  This is consistent with previously reported baryon-to-meson enhancement
  in p--Pb collisions as compared to pp collisions~\cite{Abelev:2013haa}.
  On the other hand,
  the measured $\Lambda/\Kshort$ ratio in jets is significantly smaller than
  that of inclusive distribution.
  Moreover, it is compatible with the result of the analysis applied to the
  PYTHIA-generated pp collisions. 

  \section{Conclusions}\label{sec:c04Concl}

  In contrast to the inclusive distribution,
  the $\Lambda/\Kshort$ ratio within jets in high-multiplicity p--Pb collisions
  does not exhibit baryon enhancement.
  The ratio in jets has similar magnitude and $\pT$ dependence to the ratio
  in low-multiplicity p--Pb collisions,
  and to the inclusive ratio found in pp collisions.
  It is also consistent with the analysis performed on the pp collisions
  simulated with PYTHIA8 event generator.
  It is plausible that the baryon enhancement may therefore be attributable to
  the soft (low $Q^{2}$) component of the collision as discussed
  in~\cite{Cuautle:2014yda}.
  This result disfavors the hard-soft recombination models while it is
  consistent with a picture in which the value of baryon/meson ratio has two
  independent mechanisms:
   i) the expansion of the soft particles of the underlying event within
      a common velocity field (radial flow), and
  ii) the production of particles via hard parton--parton scatterings and the
      subsequent jet fragmentation.

\bibliographystyle{LaTex/elsarticle-num}
\bibliography{V0sInJets_QM2014_XiaomingZhang}

\begin{thebibliography}{10}
\expandafter\ifx\csname url\endcsname\relax
  \def\url#1{\texttt{#1}}\fi
\expandafter\ifx\csname urlprefix\endcsname\relax\def\urlprefix{URL }\fi
\expandafter\ifx\csname href\endcsname\relax
  \def\href#1#2{#2} \def\path#1{#1}\fi

\bibitem{Aamodt:2011qy}
ALICE Collaboration, K.~Aamodt, et~al., {Centrality Dependence of the
  Charged-Particle Multiplicity Density at Midrapidity in Pb--Pb Collisions at
  $\sqrt{s_{\rm NN}}=2.76$ ~TeV}, Phys.~Rev.~Lett. 106 (2011) 032301.
\newblock \href {http://dx.doi.org/10.1103/PhysRevLett.106.032301}
  {\path{doi:10.1103/PhysRevLett.106.032301}}.

\bibitem{Abelev:2012hxa}
ALICE Collaboration, B.~Abelev, et~al., {Centrality Dependence of Charged
  Particle Production at Large Transverse Momentum in Pb--Pb Collisions at
  $\sqrt{s_{\rm NN}}=2.76$~TeV}, Phys.~Lett. B720 (2013) 52--62.
\newblock \href {http://arxiv.org/abs/1208.2711} {\path{arXiv:1208.2711}},
  \href {http://dx.doi.org/10.1016/j.physletb.2013.01.051}
  {\path{doi:10.1016/j.physletb.2013.01.051}}.

\bibitem{Abelev:2013kqa}
ALICE Collaboration, B.~Abelev, et~al., {Measurement of charged jet suppression
  in Pb--Pb collisions at $\sqrt{s_{\rm NN}}=2.76$~TeV}, JHEP 1403 (2014) 013.
\newblock \href {http://arxiv.org/abs/1311.0633} {\path{arXiv:1311.0633}},
  \href {http://dx.doi.org/10.1007/JHEP03(2014)013}
  {\path{doi:10.1007/JHEP03(2014)013}}.

\bibitem{Abelev:2014dsa}
ALICE Collaboration, B.~Abelev, et~al., {Transverse momentum dependence of
  inclusive primary charged-particle production in p--Pb collisions at
  $\sqrt{s_{\rm NN}}=5.02$~TeV}, CERN-PH-EP-2014-088 (2014).
\newblock \href {http://arxiv.org/abs/1405.2737} {\path{arXiv:1405.2737}}.

\bibitem{Abelev:2012ola}
ALICE Collaboration, B.~Abelev, et~al., {Long-range angular correlations on the
  near and away side in p--Pb collisions at $\sqrt{s_{\rm NN}}=5.02$~TeV},
  Phys.~Lett. B719 (2013) 29--41.
\newblock \href {http://arxiv.org/abs/1212.2001} {\path{arXiv:1212.2001}},
  \href {http://dx.doi.org/10.1016/j.physletb.2013.01.012}
  {\path{doi:10.1016/j.physletb.2013.01.012}}.

\bibitem{ABELEV:2013wsa}
ALICE Collaboration, B.~Abelev, et~al., {Long-range angular correlations of
  $\pi$, K and p in p--Pb collisions at $sqrt{s_{\rm NN}}=5.02$~TeV},
  Phys.~Lett. B726 (2013) 164--177.
\newblock \href {http://arxiv.org/abs/1307.3237} {\path{arXiv:1307.3237}},
  \href {http://dx.doi.org/10.1016/j.physletb.2013.08.024}
  {\path{doi:10.1016/j.physletb.2013.08.024}}.

\bibitem{Aamodt:2010uf}
ALICE Collaboration, K.~Aamodt, et~al., {Elliptic Flow of Charged Particles in
  Pb--Pb Collisions at $\sqrt{s_{\rm NN}}=2.76$~TeV}, Phy.~Rev.~Lett. 105
  (2010) 252302.
\newblock \href {http://dx.doi.org/10.1103/PhysRevLett.105.252302}
  {\path{doi:10.1103/PhysRevLett.105.252302}}.

\bibitem{Strikman:2011cx}
M.~Strikman, {Transverse Nucleon Structure and Multiparton Interactions},
  Acta~Phys.~Polon. B42 (2011) 2607--2630.
\newblock \href {http://arxiv.org/abs/1112.3834} {\path{arXiv:1112.3834}},
  \href {http://dx.doi.org/10.5506/APhysPolB.42.2607}
  {\path{doi:10.5506/APhysPolB.42.2607}}.

\bibitem{Bozek:2013uha}
P.~Bozek, W.~Broniowski, {Collective dynamics in high-energy proton--nucleus
  collisions}, Phys.~Rev. C88 (2013) 014903.
\newblock \href {http://arxiv.org/abs/1304.3044} {\path{arXiv:1304.3044}},
  \href {http://dx.doi.org/10.1103/PhysRevC.88.014903}
  {\path{doi:10.1103/PhysRevC.88.014903}}.

\bibitem{Shuryak:2013ke}
E.~Shuryak, I.~Zahed, {High-multiplicity pp and pA collisions: Hydrodynamics at
  its edge}, Phys.~Rev. C88 (2013) 044915.
\newblock \href {http://arxiv.org/abs/1301.4470} {\path{arXiv:1301.4470}},
  \href {http://dx.doi.org/10.1103/PhysRevC.88.044915}
  {\path{doi:10.1103/PhysRevC.88.044915}}.

\bibitem{Abelev:2013haa}
ALICE Collaboration, B.~Abelev, et~al., {Multiplicity Dependence of Pion, Kaon,
  Proton and Lambda Production in p--Pb Collisions at $\sqrt{s_{\rm
  NN}}=5.02$~TeV}, Phys.~Lett. B728 (2014) 25--38.
\newblock \href {http://arxiv.org/abs/1307.6796} {\path{arXiv:1307.6796}},
  \href {http://dx.doi.org/10.1016/j.physletb.2013.11.020}
  {\path{doi:10.1016/j.physletb.2013.11.020}}.

\bibitem{Abelev:2013xaa}
ALICE Collaboration, B.~Abelev, et~al., {${\rm K}_{\rm S}^{0}$ and $\Lambda$
  production in Pb--Pb collisions at $\sqrt{s_{\rm NN}}=2.76$~TeV},
  Phys.~Rev.~Lett. 111 (2013) 222301.
\newblock \href {http://arxiv.org/abs/1307.5530} {\path{arXiv:1307.5530}},
  \href {http://dx.doi.org/10.1103/PhysRevLett.111.222301}
  {\path{doi:10.1103/PhysRevLett.111.222301}}.

\bibitem{Bozek:2011gq}
P.~Bozek, {Hydrodynamic flow from RHIC to LHC}, Acta Phys.~Polon. B43 (2012)
  689.
\newblock \href {http://arxiv.org/abs/1111.4398} {\path{arXiv:1111.4398}},
  \href {http://dx.doi.org/10.5506/APhysPolB.43.689}
  {\path{doi:10.5506/APhysPolB.43.689}}.

\bibitem{Ortiz:2013yxa}
A.~Ortiz~Velasquez, P.~Christiansen, E.~Cuautle~Flores, I.~Maldonado~Cervantes,
  G.~Paice, {Color Reconnection and Flowlike Patterns in pp Collisions},
  Phys.~Rev.~Lett. 111~(4) (2013) 042001.
\newblock \href {http://arxiv.org/abs/1303.6326} {\path{arXiv:1303.6326}},
  \href {http://dx.doi.org/10.1103/PhysRevLett.111.042001}
  {\path{doi:10.1103/PhysRevLett.111.042001}}.

\bibitem{Fries:2003vb}
R.~Fries, B.~Muller, C.~Nonaka, S.~Bass, {Hadronization in heavy ion
  collisions: Recombination and fragmentation of partons}, Phys.~Rev.~Lett. 90
  (2003) 202303.
\newblock \href {http://arxiv.org/abs/nucl-th/0301087}
  {\path{arXiv:nucl-th/0301087}}, \href
  {http://dx.doi.org/10.1103/PhysRevLett.90.202303}
  {\path{doi:10.1103/PhysRevLett.90.202303}}.

\bibitem{Sjostrand:2007gs}
T.~Sjostrand, S.~Mrenna, P.~Z. Skands, {A Brief Introduction to PYTHIA 8.1},
  Comput.~Phys.~Commun. 178 (2008) 852--867.
\newblock \href {http://arxiv.org/abs/0710.3820} {\path{arXiv:0710.3820}},
  \href {http://dx.doi.org/10.1016/j.cpc.2008.01.036}
  {\path{doi:10.1016/j.cpc.2008.01.036}}.

\bibitem{Aamodt:2008zz}
ALICE Collaboration, K.~Aamodt, et~al., {The ALICE experiment at the CERN LHC},
  JINST 3 (2008) S08002.
\newblock \href {http://dx.doi.org/10.1088/1748-0221/3/08/S08002}
  {\path{doi:10.1088/1748-0221/3/08/S08002}}.

\bibitem{Abelev:2014ffa}
ALICE Collaboration, B.~Abelev, et~al., {Performance of the ALICE Experiment at
  the CERN LHC}, CERN-PH-EP-2014-031 (2014).
\newblock \href {http://arxiv.org/abs/1402.4476} {\path{arXiv:1402.4476}}.

\bibitem{Toia:2014cd}
A.~Toia for~the ALICE~Collaboration, {Centrality Dependence of Charged Particle
  Production in p--A collisions measured by ALICE}, these proceedings (2014).

\bibitem{Abelev:2013fn}
ALICE Collaboration, B.~Abelev, et~al., {Measurement of the inclusive
  differential jet cross section in pp collisions at $\sqrt{s}=2.76$~TeV},
  Phys.~Lett. B722 (2013) 262--272.
\newblock \href {http://arxiv.org/abs/1301.3475} {\path{arXiv:1301.3475}},
  \href {http://dx.doi.org/10.1016/j.physletb.2013.04.026}
  {\path{doi:10.1016/j.physletb.2013.04.026}}.

\bibitem{Cacciari:2008gp}
M.~Cacciari, G.~P. Salam, G.~Soyez, {The Anti-$k_{\rm T}$ jet clustering
  algorithm}, JHEP 0804 (2008) 063.
\newblock \href {http://arxiv.org/abs/0802.1189} {\path{arXiv:0802.1189}},
  \href {http://dx.doi.org/10.1088/1126-6708/2008/04/063}
  {\path{doi:10.1088/1126-6708/2008/04/063}}.

\bibitem{Cacciari:2007fd}
M.~Cacciari, G.~P. Salam, {Pileup subtraction using jet areas}, Phys.~Lett.
  B659 (2008) 119--126.
\newblock \href {http://arxiv.org/abs/0707.1378} {\path{arXiv:0707.1378}},
  \href {http://dx.doi.org/10.1016/j.physletb.2007.09.077}
  {\path{doi:10.1016/j.physletb.2007.09.077}}.

\bibitem{Chatrchyan:2012tt}
CMS Collaboration, S.~Chatrchyan, et~al., {Measurement of the underlying event
  activity in pp collisions at $\sqrt{s}=0.9$ and $7$~TeV with the novel
  jet-area/median approach}, JHEP 1208 (2012) 130.
\newblock \href {http://arxiv.org/abs/1207.2392} {\path{arXiv:1207.2392}},
  \href {http://dx.doi.org/10.1007/JHEP08(2012)130}
  {\path{doi:10.1007/JHEP08(2012)130}}.

\bibitem{Chinellato:2012lf}
D.~D. Chinellato, {${\rm K}_{\rm S}^{0}$, $\Lambda$ and $\bar{\Lambda}$
  production in proton--proton collisons at $\s=7$~TeV}, ALICE-ANA-501 (2012).

\bibitem{Cuautle:2014yda}
E.~Cuautle, R.~Jimenez, I.~Maldonado, A.~Ortiz, G.~Paic, et~al., {Disentangling
  the soft and hard components of the pp collisions using the sphero(i)city
  approach} (2014).
\newblock \href {http://arxiv.org/abs/1404.2372} {\path{arXiv:1404.2372}}.

\end{thebibliography}

\end{document}